\documentclass[12pt]{article}
\usepackage{epsfig}
\usepackage{rotating}
\def\br(#1,#2){\left\langle#1#2\right\rangle}
\def\sq(#1,#2){\left[#1#2\right]}
\def\s(#1,#2){s_{#1 #2}}
\def\t(#1,#2,#3){s_{#1 #2 #3}}

\begin{document}
\begin{titlepage}

\hspace*{\fill}\parbox[t]{5cm}
{hep-ph/0611348 \\
FERMILAB-Pub-06/368-T \\
CP3-06-16 \\
ILL-(TH)-06-07 \\
\today} \vskip2cm
\begin{center}
{\Large \bf Production of a $W$ Boson and Two Jets \\
\bigskip  with One $b$-quark Tag} \\
\medskip
\bigskip\bigskip\bigskip\bigskip
{\large  {\bf J.~Campbell},$^1$
         {\bf R.~K.~Ellis},$^2$
         {\bf F.~Maltoni},$^3$
     and {\bf S.~Willenbrock}$^4$} \\
\bigskip\bigskip\medskip
$^{1}$Department of Physics and Astronomy, University of Glasgow \\
Glasgow G12 8QQ, United Kingdom \\
\bigskip
$^{2}$Theoretical Physics Department, Fermi National Accelerator Laboratory \\
P.~O.~Box 500, Batavia, IL\ \ 60510 \\ \bigskip
$^{3}$Institut de Physique Th\'{e}orique and \\
Centre for Particle Physics and Phenomenology (CP3) \\
Universit\'{e} Catholique de Louvain\\[1mm]
Chemin du Cyclotron 2 \\
B-1348 Louvain-la-Neuve, Belgium \\
\bigskip
$^{4}$Department of Physics, University of Illinois at Urbana-Champaign \\
1110 West Green Street, Urbana, IL\ \ 61801 \\ \bigskip
\end{center}

\bigskip\bigskip\bigskip

\begin{abstract}
The production of a $W$ boson and two jets, at least one of which
contains a $b$ quark, is a principal background to single-top
production, Higgs production, and signals of new physics at hadron
colliders.  We present a next-to-leading-order (NLO) calculation of
the cross section at the Fermilab Tevatron and the CERN Large Hadron
Collider.  The NLO cross section differs substantially from that at
LO, and we provide a context in which to understand this result.
\end{abstract}

\end{titlepage}

\section{Introduction}
\label{sec:intro}

Many signals for new physics at hadron colliders involve an
electroweak gauge boson ($\gamma,Z,W$) and jets, one or more of
which contain a heavy quark ($c,b$).  The prime example is the
discovery of the top quark via the signal $W+4j$, where two of the
jets contain $b$ quarks \cite{Abe:1995hr,Abachi:1995iq}.  It is
important to understand the backgrounds to these signals in as much
detail as possible \cite{Acosta:2001ct}.  We have recently completed
a next-to-leading-order (NLO) calculation of $Z$ production in
association with one or two jets, one (or more) of which contain a
heavy quark
\cite{Campbell:2000bg,Campbell:2003dd,Campbell:2005zv}.\footnote{The
inclusive production of a $Z$ with one or more heavy quarks is
presented in Ref.~\cite{Maltoni:2005wd}.}  In this paper we present
a NLO calculation of $W$ production in association with two jets,
one or more of which contain a $b$ quark.  The case of $W$
production in association with one or more jets containing a charm
quark is more complicated due to the additional contributions coming
from $s\to Wc$.

The production of a $W$ boson in association with two jets, one or
more of which contain a heavy quark, is particularly interesting as
it is the principal background to both single-top production
\cite{Acosta:2004bs,Abazov:2005zz,Abazov:2006uq} and Higgs
production (via $q\bar q'\to Wh$)
\cite{Abazov:2004jy,Acosta:2005ga}, which are currently being sought
at the Fermilab Tevatron. Single-top production occurs via both a
$t$-channel process, $qb\to q't$
\cite{Willenbrock:1986cr,Yuan:1989tc,Ellis:1992yw}, and an
$s$-channel process, $q\bar q'\to t\bar b$
\cite{Cortese:1991fw,Stelzer:1995mi}. In both cases the final state,
after top decay, comprises a $W$ boson and two jets, one
($t$-channel) or both ($s$-channel) of which contain a $b$ quark. In
the case of Higgs production, the final state (after the decay $h\to
b\bar b$) comprises a $W$ and two jets, both of which contain a $b$
quark \cite{Stange:1993ya,Stange:1994bb}. However, even if both jets
contain a $b$ quark, it is more efficient to tag only one of them
\cite{Acosta:2005ga}. Thus in all cases the principal background is
$W$ plus two jets, one (or more) of which contain a heavy quark. The
signals for single top
\cite{Bordes:1994ki,Smith:1996ij,Stelzer:1997ns,Chetyrkin:2000mq,
Harris:2002md,Cao:2004ky,Cao:2004ap,Sullivan:2004ie,
Campbell:2004ch,Cao:2005pq,Frixione:2005vw} and Higgs
\cite{Han:1991ia,Brein:2003wg,Ciccolini:2003jy} are known at NLO and
beyond; our goal is to provide a calculation of the principal
background at NLO.

\begin{figure}[ht]
\begin{center}
\epsfxsize=5cm \epsfbox{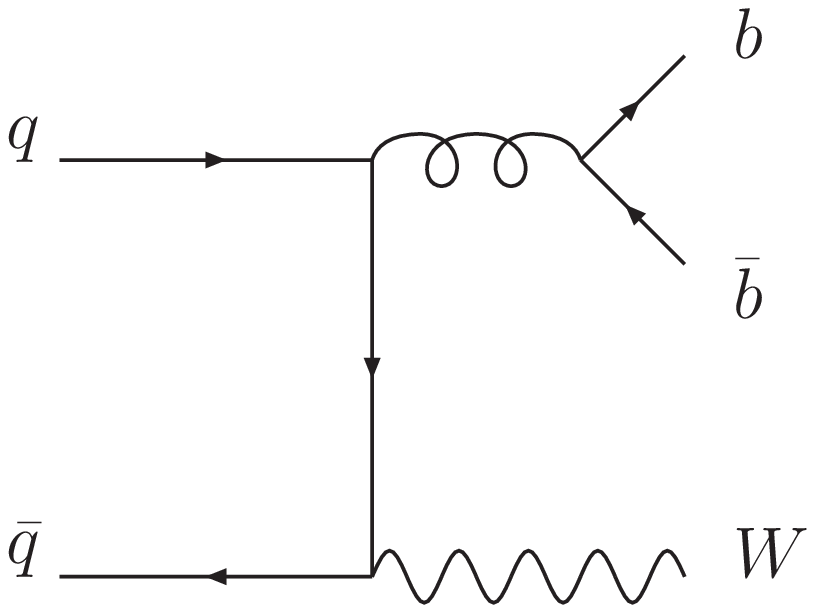} \hspace*{1cm}
\epsfxsize=5cm \epsfbox{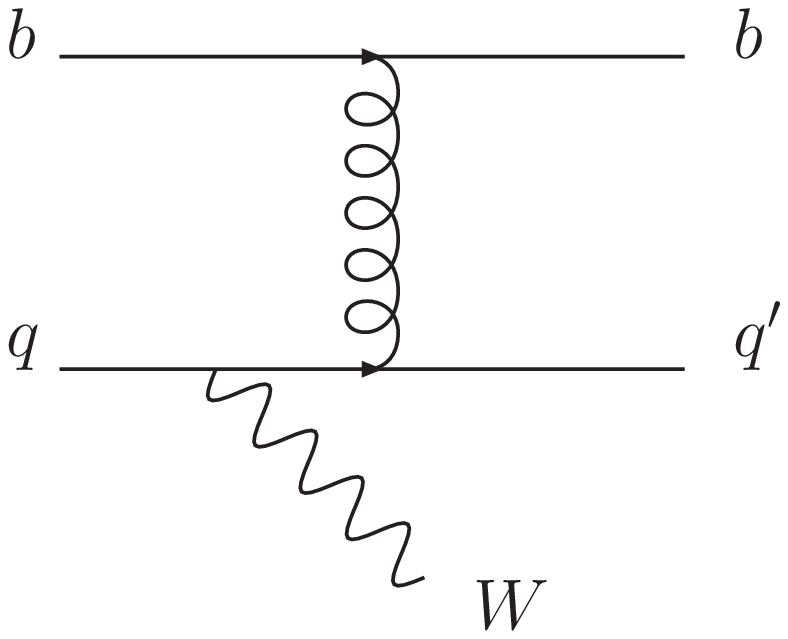} \\
(a) \hspace*{6cm} (b)
\end{center}
\caption{Leading-order processes for the production of a $W$ boson
and two jets, at least one of which contains a $b$ quark.}
\label{fig:Wbb}
\end{figure}

At leading order (LO), there are two processes that produce a $W$
boson and two jets, at least one of which contains a $b$ quark.
These are shown in Fig.~\ref{fig:Wbb}.  The process $q\bar q'\to
Wb\bar b$ yields two $b$ jets, and is already known at NLO
\cite{Ellis:1998fv,Campbell:2003hd,FebresCordero:2006sj}. In
contrast, the process $bq\to Wbq'$ yields just one $b$ jet. For a
signal with just one $b$ tag, these two processes are comparable at
the Tevatron, while at the CERN Large Hadron Collider (LHC) the
latter process is dominant \cite{Mangano:2001xp}.

\begin{figure}[ht]
\begin{center}
\vspace*{.2cm} \hspace*{0cm} \epsfxsize=7cm \epsfbox{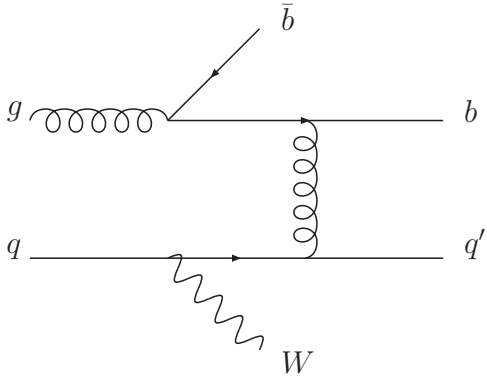}
\vspace*{-.8cm}
\end{center}
\caption{An alternative way of calculating the process in
Fig.~\ref{fig:Wbb}(b).} \label{fig:Wbbj}
\end{figure}

The process $bq\to Wbq'$ requires further consideration, as it
contains a $b$ quark in the initial state.  The $b$ distribution in
the proton is derived perturbatively via the
Dokshitzer-Gribov-Lipatov-Altarelli-Parisi (DGLAP) evolution
equations \cite{Aivazis:1993pi,Collins:1998rz}. Alternatively, one
may eschew a $b$ distribution function, and use a gluon in the
initial state, which then splits to $b\bar b$, with one $b$
participating in the hard scattering while the other remains at low
transverse momentum; this is shown in Fig.~\ref{fig:Wbbj}
\cite{Mangano:2001xp}. However, there are two advantages to working
with a $b$ distribution function. First, initial-state collinear
logarithms, of order $[\alpha_S\ln(M_W/m_b)]^n$, are summed to all
orders, yielding a more convergent perturbative expansion.  Second,
the LO process is simpler, which makes the NLO calculation
tractable.

A $b$ distribution function is useful when the scale of the process
is much greater than the $b$ mass.  In beauty production at HERA,
the scale is not much greater than the $b$ mass, and the appropriate
description is $\gamma^*g\to b\bar b$ \cite{Grab:2006qb}.  Thus
there is no direct measurement of the $b$ distribution function at
this time.  Single-top production ($t$-channel) may provide the
first such measurement.

The paper is organized as follows.  In Section~\ref{sec:Wbj} we
outline the NLO calculation and discuss some of the finer points.
The issue of correctly treating the $b$-quark mass arises in a novel
way, and we discuss this in some detail.  The results of the
calculation are presented in Section~\ref{sec:results}.

\section{$Wbj$ at NLO}\label{sec:Wbj}

The LO cross sections for $Wbj$ and $Wb\bar b$, arising from the
processes in Figs.~\ref{fig:Wbb}(b) and \ref{fig:Wbb}(a),
respectively, are given in parentheses in
Table~\ref{tab:Wbj_exclusive}.  All jets are required to have
$p_T>15$ GeV and $|\eta|<2$ at the Tevatron, while at the LHC we
require $p_T>25$ GeV and $|\eta|<2.5$.  The jets are required to be
separated by $\Delta R_{jj}>0.7$.  As mentioned in the Introduction,
these cross sections are comparable at the Tevatron, while $Wbj$ is
dominant at the LHC. This qualitative result is already well known
from the LO calculation of Ref.~\cite{Mangano:2001xp}, which used a
gluon in the initial state (Fig.~\ref{fig:Wbbj}) rather than a $b$
quark [Fig.~\ref{fig:Wbb}(b)]. Our goal is to calculate these cross
sections at NLO.

The processes involved in the calculation are as follows:
\begin{itemize}
\item $q\bar q'\to Wb\bar b$ at tree level [Fig.~\ref{fig:Wbb}(a)] and one loop
\item $bq\to Wbq'$ at tree level [Fig.~\ref{fig:Wbb}(b)] and one loop
\item $q\bar q'\to Wb\bar bg$ at tree level
\item $bq\to Wbq'g$ at tree level
\item $gq\to Wb\bar bq'$ at tree level (Fig.~\ref{fig:Wbbj})
\item $bg\to Wbq\bar q'$ at tree level
\end{itemize}
In all cases $q=u,d,s,c$, and we include Cabbibo-Kobayashi-Maskawa
(CKM) mixing.  The $b$-quark mass is neglected throughout (except
where noted).  The calculation is simpler than the corresponding
calculation for $ZQj$ and $ZQ\overline Q$ because there are several
subprocesses that have no contributing analogue, namely $Qg\to ZQg$
and $gg\to ZQ\overline Q$ \cite{Campbell:2005zv}. The analogous
processes, $bg\to Wtg$ and $gg\to Wt\bar b$, yield a top quark in
the final state.  We do not consider these, or any other processes
that yield a top quark, as part of our calculation of the $Wbj$ and
$Wb\bar b$ backgrounds.

The NLO calculations in this paper were performed with the
Monte-Carlo code MCFM \cite{Ellis:1999ec,MCFM}.  The leading-order
calculations were performed both with this code and with MadEvent
\cite{Maltoni:2002qb}.  The NLO corrections are included in the code
MCFM by implementing the virtual helicity amplitudes of
Ref.~\cite{Bern:1997sc}. These are given in the four-dimensional
helicity scheme, which is used throughout the calculation.  The
strong coupling constant is shifted from this scheme to the
$\overline{\rm MS}$ scheme \cite{Catani:1996pk}.  The real
corrections are adapted from Ref.~\cite{Nagy:1998bb} and
singularities are handled using the dipole subtraction
method~\cite{Catani:1996vz}. Since the matrix elements contain the
decay of the gauge boson into massless particles, the code is
general enough to provide results for final states from $W^*\to
\ell\nu$, including the correlation between the $W^*$ spin and the
angular distribution of the leptons. For this paper we specialize to
the case of an on-shell $W$ boson.

The NLO cross sections are presented in
Table~\ref{tab:Wbj_exclusive}.  To obtain the NLO cross sections for
$Wbj$ and $Wb\bar b$, which involve the radiation of additional
partons, two partons are combined into a single jet by adding their
four-momenta if $\Delta R_{jj}<0.7$.  If the combined partons are
both $b$ quarks, then the process contributes to the column labeled
$W(b\bar b)j$. This is a $W+2j$ event in which one jet contains two
$b$ quarks, which changes the tagging probability for that jet
\cite{Acosta:2004nj}. It is calculated with a finite $b$-quark mass
($m_b=4.75$ GeV) in order to regulate the logarithmic divergence
present when the heavy quarks are collinear. If all three partons
are well separated, then the process contributes to either $Wb\bar
bj$ or $Wbjj$.

There is another aspect of the calculation in which the $b$-quark
mass cannot be neglected.  If the $b$ quark comes from a final-state
virtual gluon splitting to $b\bar b$, with the other $b$ quark
missed, then the cross section is sensitive to the $b$-quark mass.
This can occur in the NLO processes $q\bar q'\to Wb\bar bg$ and
$gq\to Wb\bar bq'$, both of which contribute to $Wbj$ when one $b$
quark is missed.  We therefore calculated these processes with a
finite $b$-quark mass. In the case of the process $gq\to Wb\bar
bq'$, which involves the splitting $g\to b\bar b$ in the initial
state (see Fig.~\ref{fig:Wbbj}) as well as the final state, this
requires that we abandon the dipole subtraction method in favor of
subtracting the mass singularity via the truncated $b$ distribution
function \cite{Aivazis:1993pi,Collins:1998rz}
\begin{equation}
\tilde b(x,\mu) =
\frac{\alpha_S(\mu)}{2\pi}\ln\left(\frac{\mu^2}{m_b^2}\right)\int_x^1\frac{dy}{y}P_{qg}\left(\frac{x}{y}\right)g(y,\mu)
\end{equation}
where $P_{qg}(z)={1\over 2}[z^2+(1-z)^2]$ is the DGLAP splitting
function.  The counterterm is constructed by calculating $\tilde
bq\to Wbq'$; this cancels the initial-state logarithmic dependence
on the $b$-quark mass in $gq\to Wb\bar bq'$, and yields a cross
section in the $\overline{\rm MS}$ factorization scheme.

It is interesting to compare the LO result for $Wbj$ with a
calculation based on $gq\to Wq'b\bar b$ (Fig.~\ref{fig:Wbbj})
\cite{Mangano:2001xp}, which does not use a $b$ distribution
function.  With the same parameters as
Table~\ref{tab:Wbj_exclusive}, and using $m_b=4.75$ GeV and CTEQ6L1
parton distribution functions \cite{Pumplin:2002vw}, we find a cross
section of 1.92 pb at the Tevatron.  This is to be compared with
1.06 pb using $bq\to Wbq'$, and serves as a rough check of the
formalism. Both of these numbers are LO and very scale dependent, so
one should expect only qualitative agreement.\footnote{Including the
uncertainties due to the independent variation of the
renormalization and factorization scales from $M_W/2$ to $2M_W$, the
cross sections are $1.92^{+0.84}_{-0.51}\,^{+0.22}_{-0.15}$ pb for
$gq\to Wq'b\bar b$ and $1.06^{+0.28}_{-0.20}\,^{+0.03}_{-0.07}$ pb
for $bq\to Wbq'$} The result at the LHC ($W^++W^-$) is 81 pb, which
is to be compared with the 87.3 pb result obtained using $bq\to
Wbq'$ (see Table~\ref{tab:Wbj_exclusive}).  In any case, the NLO
results presented in this paper should be more accurate than any of
these LO results.

We checked that the effect of the heavy-quark mass is small if all
$b$ quarks are at high $p_T$ by comparing $Wb\bar b$ at tree level
with and without a finite quark mass. With the cuts used in this
paper, we find a reduction of the $Wb\bar b$ cross section by about
7\% at the Tevatron ($p_T> 15$ GeV) and 2\% at the LHC ($p_T> 25$
GeV). This is the same as the result found for the effect of the
$b$-quark mass on the full NLO calculation at the Tevatron
\cite{FebresCordero:2006sj}, which suggests that the effect of the
$b$-quark mass can be divined from a LO calculation. We found that
the heavy-quark mass is a similarly small effect for $Wb\bar bj$ at
tree level, as expected.

We also list in Table~\ref{tab:Wbj_exclusive} the LO (in
parentheses) and NLO cross sections for $Wjj$
\cite{Campbell:2003hd,Campbell:2002tg}, and the LO cross section for
$Wjjj$.  We ignore CKM mixing, which is a negligible effect at LO,
and is therefore also negligible at NLO. In these cross sections we
have included the contribution from light partons as well as heavy
quarks ($c,b$). Thus, for example, the fraction of $W+2j$ events in
which only one of the jets can be tagged as $b$ jet is given by
$[Wbj+W(b\bar b)j]/Wjj$.

\section{Results}\label{sec:results}

The results in Table~\ref{tab:Wbj_exclusive} are remarkable.  First
we see that, at NLO, the cross sections for $Wbj$ and $Wb\bar b$ are
comparable at the Tevatron.  Thus both final states are important
backgrounds to single top and Higgs.  At the LHC, $Wbj$ is much
larger than $Wb\bar b$.  At both machines, the correction to $Wbj$
is large (for $\mu_F=\mu_R=M_W$), about a factor of 1.9 at the LHC
and a factor of 2.4 at the Tevatron.  The correction is larger at
the Tevatron because $Wb\bar b$, which is comparable in size, feeds
into the $Wbj$ column at NLO when one of the $b$ quarks is outside
the fiducial region ($p_T>15$ GeV, $|\eta|<2$).

The correction to $Wb\bar b$ is significant at both machines, but
more modest than that of $Wbj$, again for $\mu_F=\mu_R=M_W$.  At the
LHC the cross section for $Wb\bar bj$ is comparable to that of
$Wb\bar b$, which calls into question the reliability of
perturbation theory. However, $Wb\bar bj$ is also a correction to
$Wbj$, and when compared with that LO cross section it is a modest
correction. Nevertheless, it does mean that when demanding two $b$
quarks in the final state, the processes $Wb\bar b$ and $Wb\bar bj$
are comparable at the LHC.  The latter is only known at LO and
therefore has a large uncertainty.  This is reflected by the large
scale dependence associated with this process, as discussed in
Refs.~\cite{Campbell:2003hd,FebresCordero:2006sj}.

Using the results in Table~\ref{tab:Wbj_exclusive}, we construct a
set of inclusive cross sections, presented in
Table~\ref{tab:Wbj_inclusive}.  Displayed this way, the radiative
corrections to the cross sections are even more significant,
especially for $Wb\bar b+X$ at the LHC, where the NLO cross section
is about 2.7 times the LO cross section.  However, the proper way to
regard this is that a new process enters at NLO, namely $gq\to
Wb\bar bq'$ (see Fig.~\ref{fig:Wbbj}), which is also a correction to
the much larger LO process $bq\to Wbq'$, as discussed in the
previous paragraph. We also list the uncertainties from varying the
renormalization scale from $M_W/2$ to $2M_W$ while keeping the
factorization scale fixed at $M_W$, from varying the factorization
scale from $M_W/2$ to $2M_W$ while keeping the renormalization scale
fixed at $M_W$, and from varying the parton distribution functions
\cite{Pumplin:2002vw}. The largest uncertainty is from varying the
renormalization scale, since the process under consideration is of
order $\alpha_S^2$ at leading order.

\begin{table}[t]
\caption{Exclusive cross sections (pb) for $W$ boson plus two (or
more) jets, with at least one $b$ jet, at the Tevatron
($\sqrt{s}=1.96$ TeV $p\bar p$, $p_T>15$ GeV and $|\eta|<2$) and LHC
($\sqrt{s}=14$ TeV $pp$, $p_T>25$ GeV and $|\eta|<2.5$). Two
final-state partons are merged into a single jet if $\Delta
R_{jj}<0.7$. No branching ratios or tagging efficiencies are
included. The labels on the columns have the following meaning:
$Wbj=$ exactly two jets, one of which contains a $b$ quark; $Wb\bar
b=$ exactly two jets, both of which contain a $b$ quark; $W(b\bar
b)j=$ exactly two jets, one of which contains two $b$ quarks;
$Wbjj=$ exactly three jets, one of which contains a $b$ quark;
$Wb\overline bj=$ exactly three jets, two of which contain a $b$
quark.  For the last set of processes, which include both light and
heavy partons in the final state, the labels mean: $Wjj=$ exactly
two jets; $Wjjj=$ exactly three jets. For $Wbj$, $Wb\bar b$, and
$Wjj$, both the leading-order (in parentheses) and
next-to-leading-order cross sections are given. The CTEQ6M parton
distribution functions are used throughout, except for the LO cross
sections in parentheses, where CTEQ6L1 is used
\cite{Pumplin:2002vw}. The factorization and renormalization scales
are chosen as $\mu_F=\mu_R=M_W$.
}
\addtolength{\arraycolsep}{0.2cm}
\renewcommand{\arraystretch}{1.5}
\medskip
\begin{center}
\begin{tabular}[5]{|c|ccc|cc|}
\hline \hline \multicolumn{1}{|c|}{} & \multicolumn{5}{c|}{Cross sections (pb) }\\[1pt]
\hline
       Collider           &  $Wbj$        & $Wb\overline b$ & $W(b\overline b)j$  &  $Wbjj$   &    $Wb\overline bj$ \\
\hline
TeV $W^+(=W^-)$           & (1.06) 2.54   & (2.48) 3.14     &     0.89            &  0.18     &    0.65 \\
LHC $W^+$                 & (51.7) 96.2  & (9.5) 14.3      &     27.0            &  13.8     &   11.6  \\
LHC $W^-$                 & (35.6) 66.4   & (6.6) 9.6       &     19.0            &  9.3     &    7.6  \\
\hline
& \multicolumn{3}{|c}{$Wjj$} & \multicolumn{2}{|c|}{$Wjjj$} \\
\hline
TeV $W^+(=W^-)$ & \multicolumn{3}{|c}{(261) 290  } &  \multicolumn{2}{|c|}{39 }  \\
LHC $W^+      $ & \multicolumn{3}{|c}{(4990) 4170  } &  \multicolumn{2}{|c|}{1280} \\
LHC $W^-      $ & \multicolumn{3}{|c}{(3650) 3030  } &  \multicolumn{2}{|c|}{890} \\
\hline
\end{tabular}
\end{center}
\label{tab:Wbj_exclusive}
\end{table}

\begin{table}[t]
\caption{Inclusive cross sections (pb) for $W$ boson plus two (or
more) jets, with at least one $b$ jet, at the Tevatron
($\sqrt{s}=1.96$ TeV $p\bar p$, $p_T>15$ GeV and $|\eta|<2$) and LHC
($\sqrt{s}=14$ TeV $pp$, $p_T>25$ GeV and $|\eta|<2.5$). Two
final-state partons are merged into a single jet if $\Delta
R_{jj}<0.7$. No branching ratios or tagging efficiencies are
included. The labels on the columns have the following meaning:
$Wbj+X=$ at least two jets, one of which contains a $b$ quark;
$Wb\bar b+X=$ at least two jets, two of which contain a $b$ quark;
$W(b\bar b)j=$ two jets, one of which contains two $b$ quarks. The
label $Wjj+X=$ at least two jets, containing both light and heavy
partons. For $Wbj+X$, $Wb\bar b+X$, and $Wjj+X$, both the
leading-order (in parentheses) and next-to-leading-order cross
sections are given. The CTEQ6M parton distribution functions are
used throughout, except for the LO cross sections in parentheses,
where CTEQ6L1 is used \cite{Pumplin:2002vw}. The factorization and
renormalization scales are chosen as $\mu_F=\mu_R=M_W$. The
uncertainties are from the variation of the renormalization scale,
the factorization scale, and the parton distribution functions,
respectively.} \addtolength{\arraycolsep}{0.2cm}
\renewcommand{\arraystretch}{1.5}
\medskip
\begin{center}
\begin{tabular}[5]{|c|ccc|}
\hline \hline \multicolumn{1}{|c|}{} & \multicolumn{3}{c|}{Cross sections (pb) }\\[1pt]
\hline
       Collider &  $Wbj+X$        & $Wb\overline b+X$ & $W(b\overline b)j$ \\
\hline
TeV $W^+(=W^-)$ &(1.06) $2.72^{+0.68}_{-0.50}\,^{+0.01}_{-0.04}\,^{+0.09}_{-0.09}$&(2.49) $3.79^{+0.56}_{-0.47}\,^{+0.12}_{-0.08}\,^{+0.11}_{-0.11}$&$0.89^{+0.36}_{-0.23}\,^{+0.10}_{-0.07}\,^{+0.03}_{-0.03}$\\
LHC $W^+$       &(51.7) $110^{+23}_{-18}\,^{+11}_{-17}\,^{+3.6}_{-3.6}$&(9.5) $25.9^{+6.2}_{-5.1}\,^{+0}_{-0.3}\,^{+0.6}_{-0.6}$&$27.0^{+11}_{-7.1}\,^{+0.1}_{-0.3}\,^{+0.9}_{-0.9}$\\
LHC $W^-$       &(35.6) $75.7^{+17}_{-12}\,^{+7.3}_{-9.5}\,^{+2.6}_{-2.6}$&(6.6) $17.2^{+4.2}_{-3.3}\,^{+0.2}_{-0.5}\,^{+0.5}_{-0.5}$&$19.0^{+7.7}_{-5.0}\,^{+0.1}_{-0.3}\,^{+0.7}_{-0.7}$\\
\hline
& \multicolumn{3}{|c|}{$Wjj+X$} \\
\hline
TeV $W^+(=W^-)$ & \multicolumn{3}{|c|}{(261) $329^{+30}_{-32}\,^{+6}_{-6}\,^{+7}_{-7}$  } \\
LHC $W^+      $ & \multicolumn{3}{|c|}{(4990) $5450^{+410}_{-480}\,^{+70}_{-0}\,^{+190}_{-190}$  } \\
LHC $W^-      $ & \multicolumn{3}{|c|}{(3650) $3920^{+300}_{-310}\,^{+60}_{-10}\,^{+150}_{-150}$  } \\
\hline
\end{tabular}
\end{center}
\label{tab:Wbj_inclusive}
\end{table}

\begin{figure}[ht]
\begin{center}
\begin{sideways}
\vspace*{.2cm} \hspace*{0cm} \epsfxsize=8cm
\epsfbox{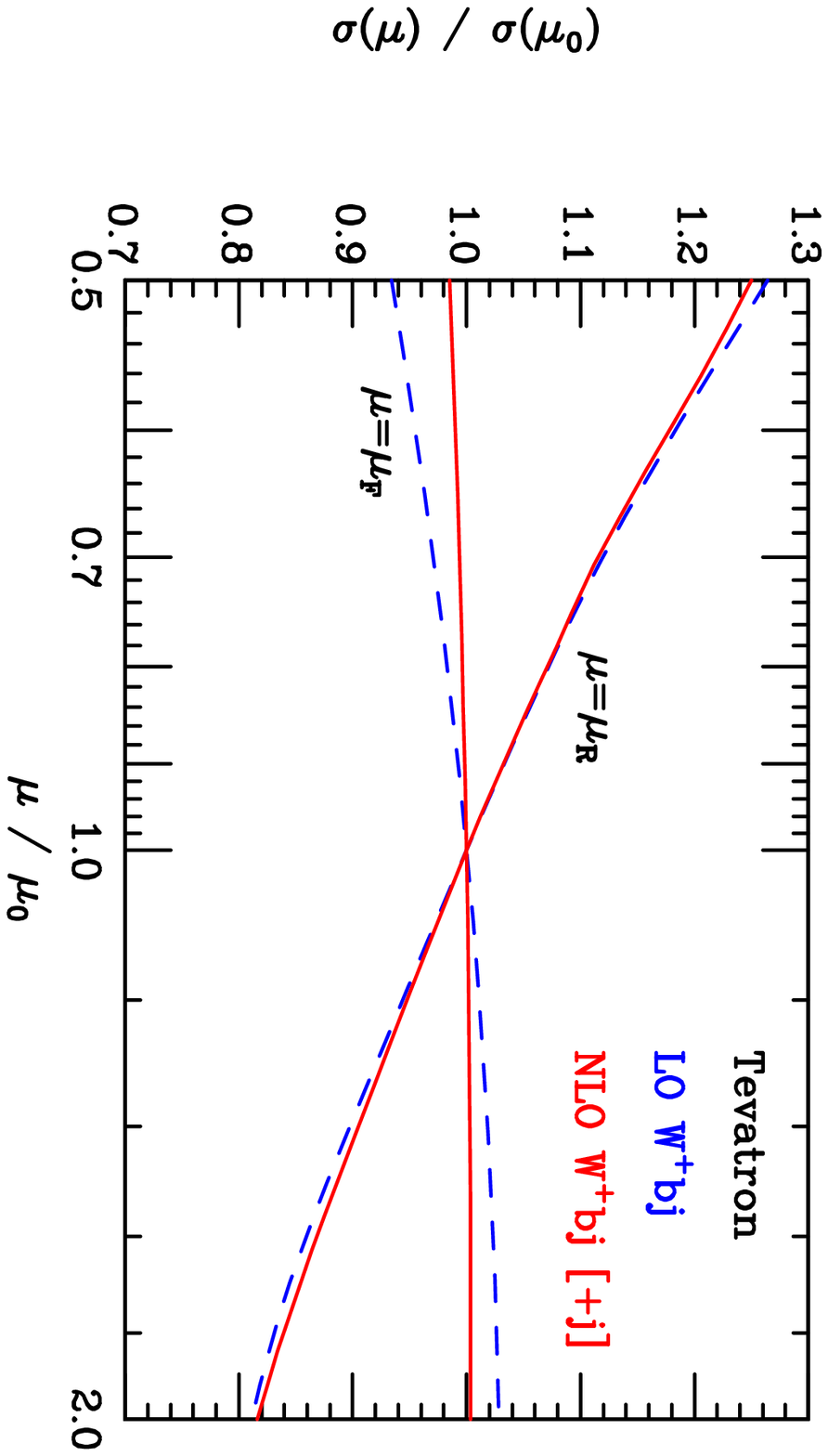}
\end{sideways}
\vspace*{-.8cm}
\end{center}
\caption{Renormalization- and factorization-scale dependence of the
inclusive $Wbj+X$ cross section at LO (dashed) and NLO (solid) at
the Tevatron. The cross section is normalized to its value at the
reference scale $\mu_0=M_W$.} \label{fig:mudep_wbj_tev}
\end{figure}

\begin{figure}[ht]
\begin{center}
\begin{sideways}
\vspace*{.2cm} \hspace*{0cm} \epsfxsize=8cm
\epsfbox{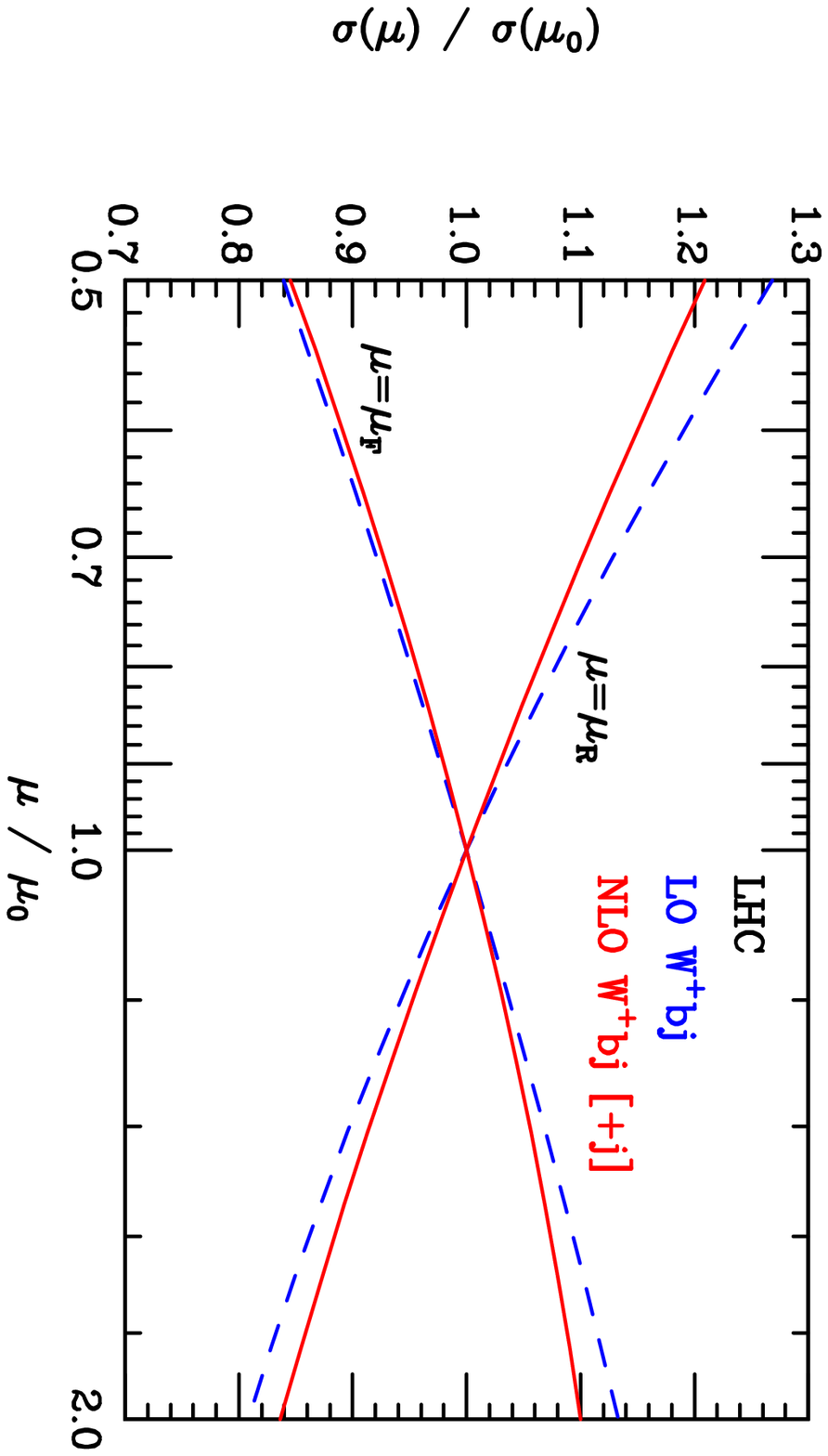}
\end{sideways}
\vspace*{-.8cm}
\end{center}
\caption{Same as Fig.~\ref{fig:mudep_wbj_tev}, but at the LHC.}
\label{fig:mudep_wbj_lhc}
\end{figure}

We show in Figs.~\ref{fig:mudep_wbj_tev} and \ref{fig:mudep_wbj_lhc}
the renormalization- and factorization-scale dependence at LO and
NLO for the inclusive $Wbj+X$ cross section at the Tevatron and the
LHC, respectively.  At the Tevatron, there is almost no reduction of
the renormalization-scale dependence at NLO.  This supports our
earlier argument that part of the source of the large NLO correction
to this process is that the process $Wb\bar bj$ contributes when one
of the $b$ jets is missed.  Since this is a tree-level process, it
has a large renormalization-scale dependence.  There is only a mild
reduction of the renormalization-scale dependence at the LHC.  At
both machines there is a mild reduction of the factorization-scale
dependence.

\begin{figure}[ht]
\begin{center}
\begin{sideways}
\vspace*{.2cm} \hspace*{0cm} \epsfxsize=11cm \epsfbox{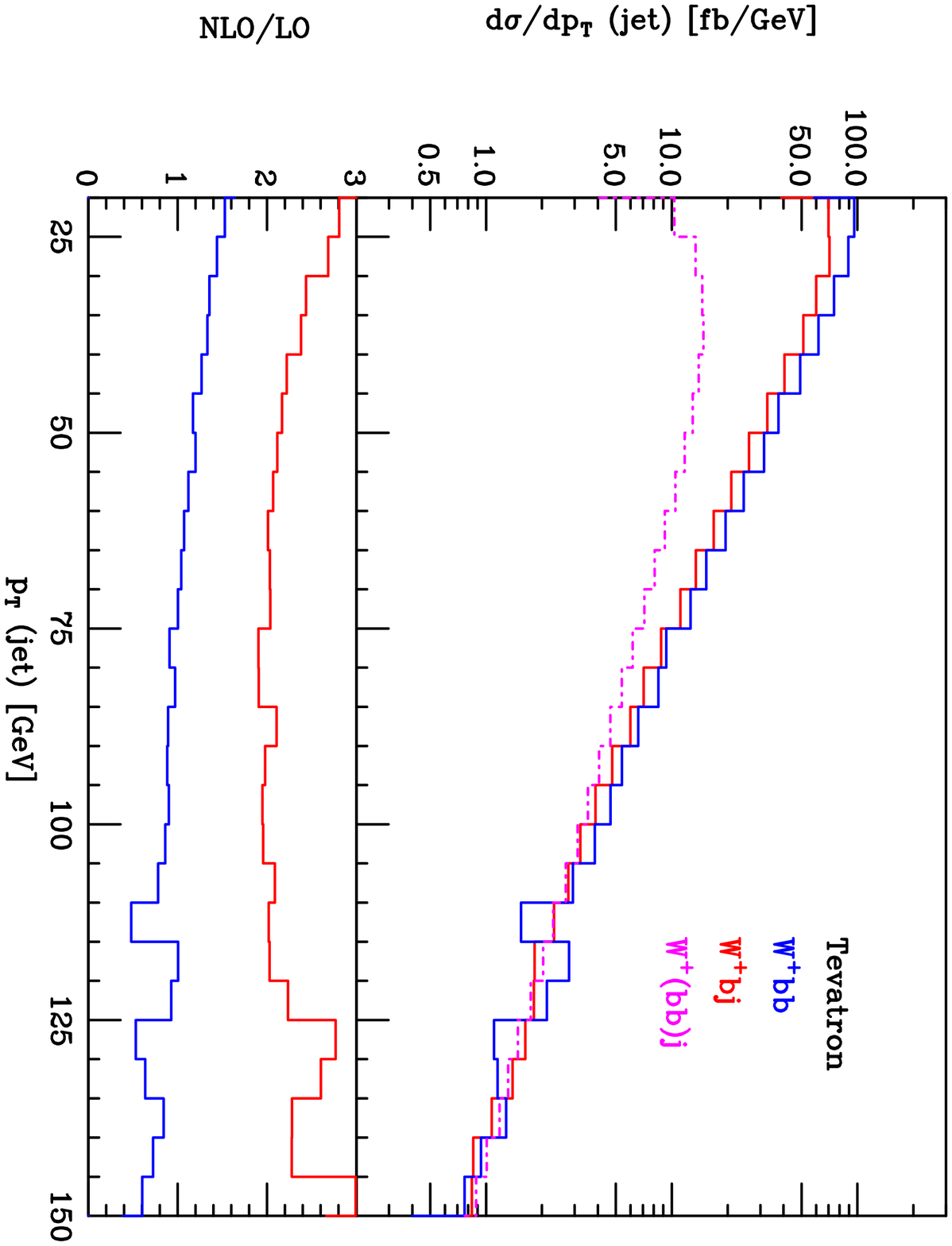}
\end{sideways}
\vspace*{-.8cm}
\end{center}
\caption{Exclusive NLO differential cross sections at the Tevatron
($\sqrt{s}=1.96$ TeV $p\bar p$, $p_T>15$ GeV, $|\eta|<2$, $\Delta
R_{jj}>0.7$) for $Wbb$ (blue or black) and $Wbj$ (red or gray),
versus the transverse momentum ($p_T$) of the highest-$p_T$ jet.
Also shown is the distribution for the NLO process $W(b\bar b)j$
(magenta or light gray, dot-dashed), in which a jet contains two $b$
quarks.  The lower histogram shows the ratio of the NLO and LO cross
sections.} \label{fig:highpt-tev}
\end{figure}

Using the code MCFM \cite{Ellis:1999ec,MCFM}, we are able to plot
any desired distribution at both LO and NLO.  For example, we show
in Fig.~\ref{fig:highpt-tev} the exclusive NLO differential cross
section at the Tevatron for $Wb\bar b$, $Wbj$, and $W(b\bar b)j$,
versus the transverse momentum ($p_T$) of the jet with the highest
$p_T$.  For $Wbj$, the highest-$p_T$ jet contains a $b$ quark $48\%$
of the time; for $W(b\bar b)j$, the percentage is $63\%$.  The
$W(b\bar b)j$ cross section, which first arises at NLO, has a very
different shape from the other two, and is only significant at high
$p_T$.

The lower histogram in Fig.~\ref{fig:highpt-tev} shows the ratio of
the NLO and LO cross sections.  We see that there is a change in
shape of the $Wbj$ and $Wb\bar b$ differential cross sections at
NLO, with relatively more events at low $p_T$. There is also an
enhancement of the $Wbj$ cross section at high $p_T$, due to the
process $q\bar q\to Wb\bar bj$ with one $b$ missed.

\begin{figure}[ht]
\begin{center}
\begin{sideways}
\vspace*{.2cm} \hspace*{0cm} \epsfxsize=11cm \epsfbox{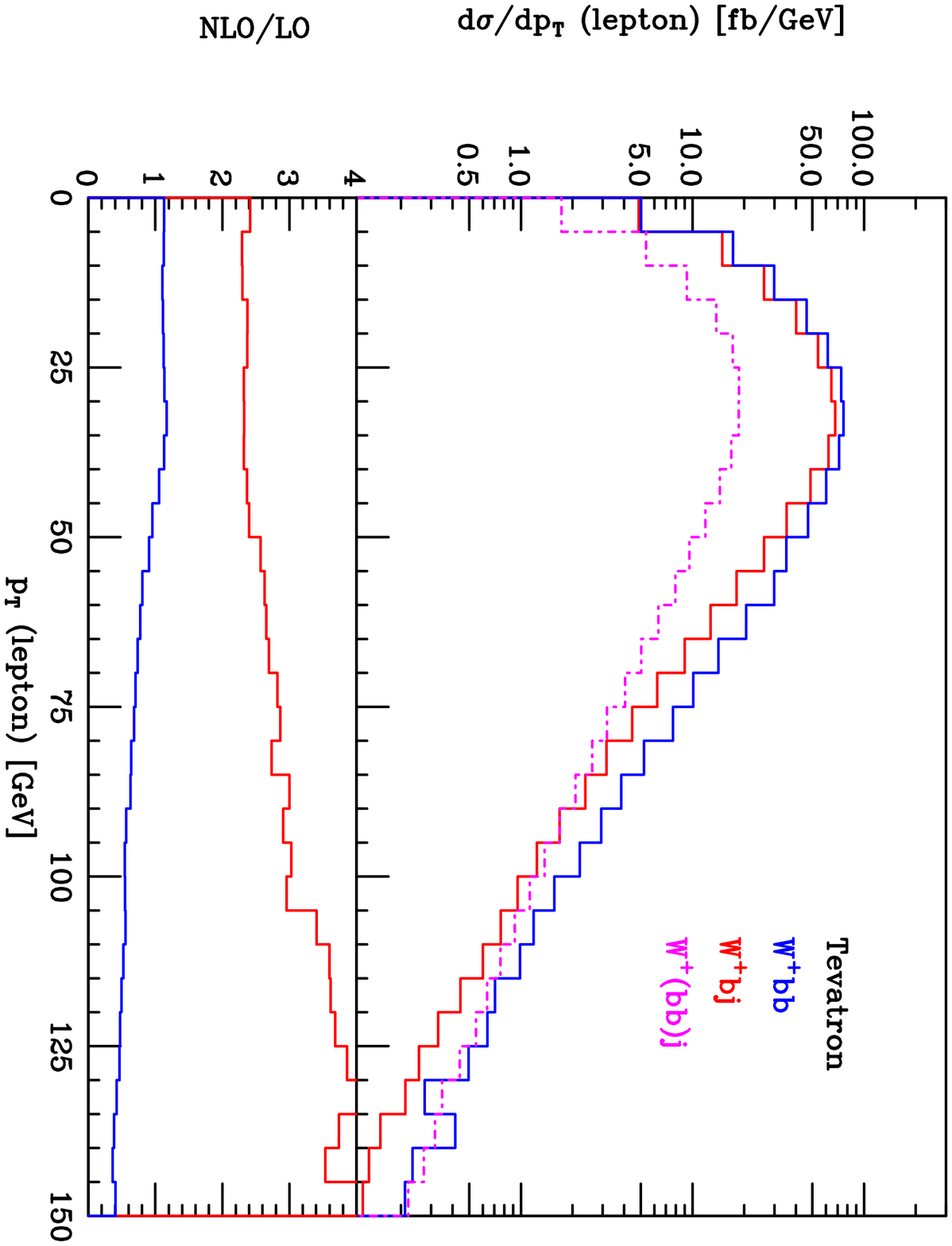}
\end{sideways}
\vspace*{-.8cm}
\end{center}
\caption{Same as Fig.~\ref{fig:highpt-tev}, but for the charged
lepton from $W$ decay.} \label{fig:lepton-tev}
\end{figure}

We show in Fig.~\ref{fig:lepton-tev} the exclusive NLO transverse
momentum distribution of the charged lepton from $W$ decay at the
Tevatron. We again see a change in shape at NLO, with relatively
more events at low $p_T$ for $Wbb$, and relatively more events at
high $p_T$ for $Wbj$, the latter again due to the process $q\bar
q\to Wb\bar bj$ with one $b$ missed. This plot demonstrates the
ability of MCFM to calculate quantities involving the decay products
of the $W$ boson while correctly treating the correlation between
the angular distribution of the decay products and the $W$ spin. The
$W$ boson is treated as on-shell in these plots, but MCFM is
flexible enough to allow for off-shell $W$ bosons as well.

\begin{figure}[ht]
\begin{center}
\begin{sideways}
\vspace*{.2cm} \hspace*{0cm} \epsfxsize=11cm
\epsfbox{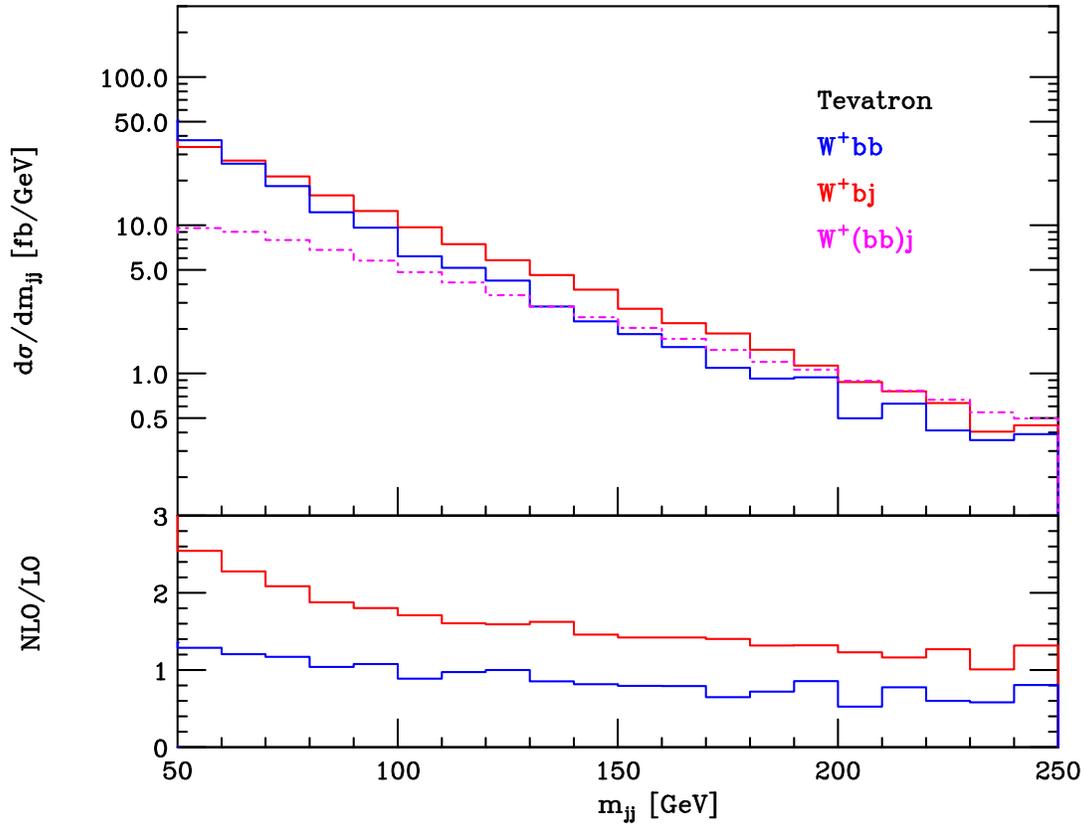}
\end{sideways}
\vspace*{-.8cm}
\end{center}
\caption{Same as Fig.~\ref{fig:highpt-tev}, but for the di-jet
invariant mass.} \label{fig:massinv-tev}
\end{figure}

We show in Fig.~\ref{fig:massinv-tev} the exclusive NLO di-jet
invariant mass at the Tevatron.  Both $Wb\bar b$ and $Wbj$ are
enhanced at low invariant masses at NLO.  The process $W(b\bar b)j$
is significant at high invariant mass.  All three processes are
numerically important as backgrounds to the search for a Higgs boson
via the process $q\bar q\to Wh\to Wb\bar b$.  For example, the
integrated cross sections in the region 110 GeV $<m_{jj}<130$ GeV
are 130 fb for $Wbj$, 95 fb for $Wb\bar b$, and 75 fb for $W(b\bar
b)j$.

\begin{figure}[ht]
\begin{center}
\begin{sideways}
\vspace*{.2cm} \hspace*{0cm} \epsfxsize=11cm \epsfbox{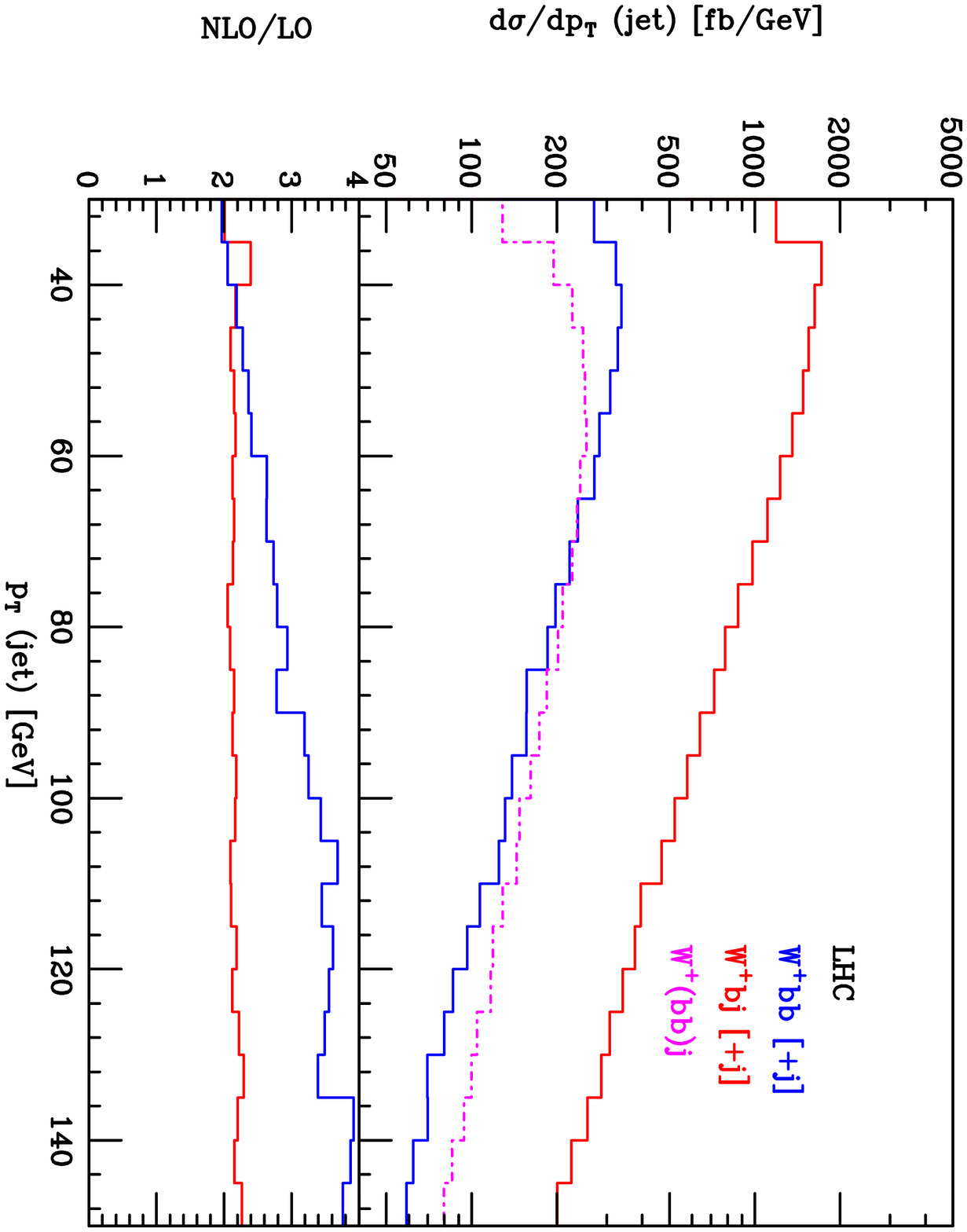}
\end{sideways}
\vspace*{-.8cm}
\end{center}
\caption{Inclusive NLO differential cross sections at the LHC
($\sqrt{s}=14$ TeV $pp$, $p_T>25$ GeV, $|\eta|<2.5$, $\Delta
R_{jj}>0.7$) for $Wbb[+j]$ (blue or black) and $Wbj[+j]$ (red or
gray), versus the transverse momentum ($p_T$) of the highest-$p_T$
jet. Also shown is the distribution for the NLO process $W(b\bar
b)j$ (magenta or light gray, dot-dashed), in which a jet contains
two $b$ quarks.  The lower histogram shows the ratio of the NLO and
LO cross sections.} \label{fig:highpt-lhc}
\end{figure}

\begin{figure}[ht]
\begin{center}
\begin{sideways}
\vspace*{.2cm} \hspace*{0cm} \epsfxsize=11cm \epsfbox{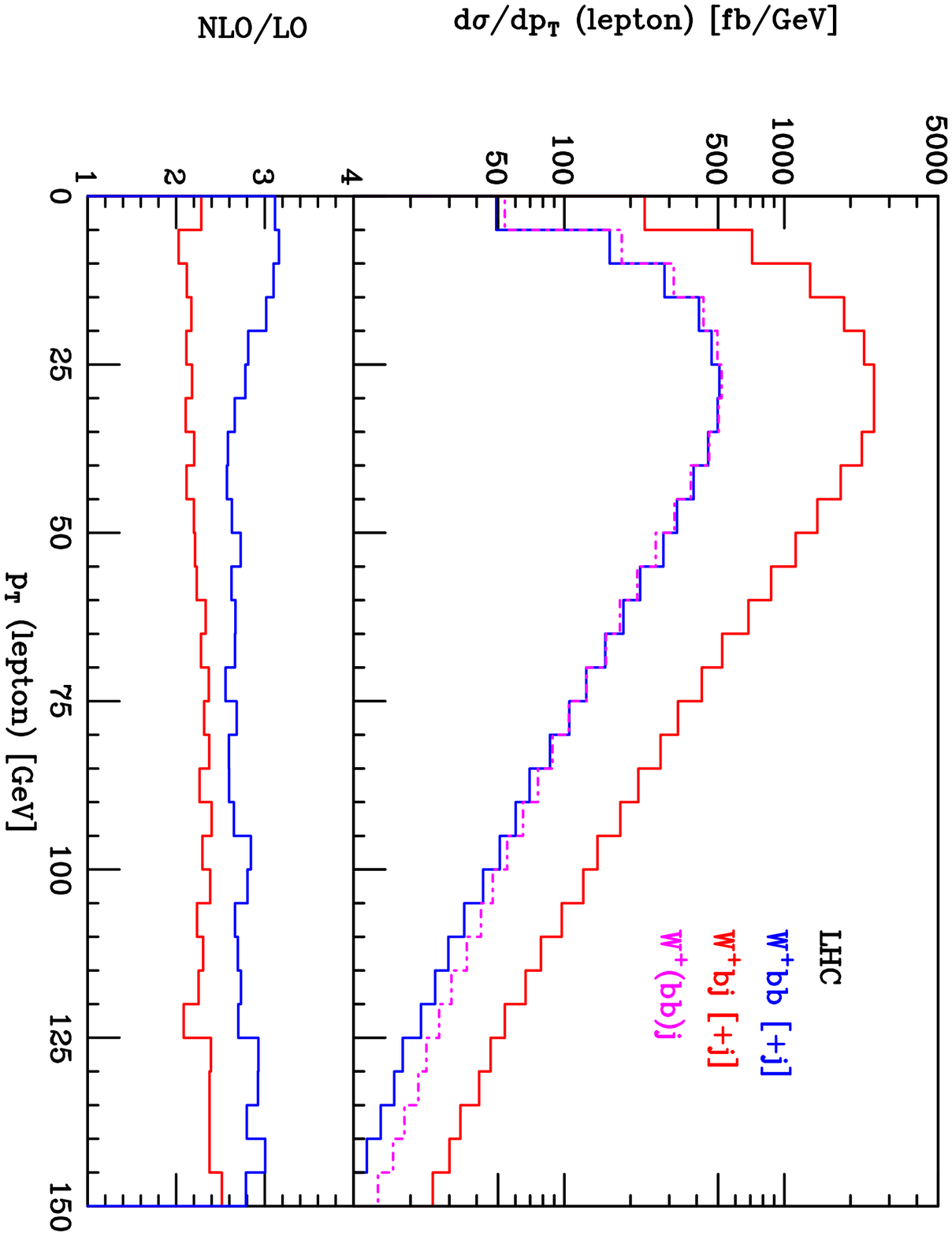}
\end{sideways}
\vspace*{-.8cm}
\end{center}
\caption{Same as Fig.~\ref{fig:highpt-lhc}, but for the charged
lepton from $W$ decay.} \label{fig:lepton-lhc}
\end{figure}

\begin{figure}[ht]
\begin{center}
\begin{sideways}
\vspace*{.2cm} \hspace*{0cm} \epsfxsize=11cm
\epsfbox{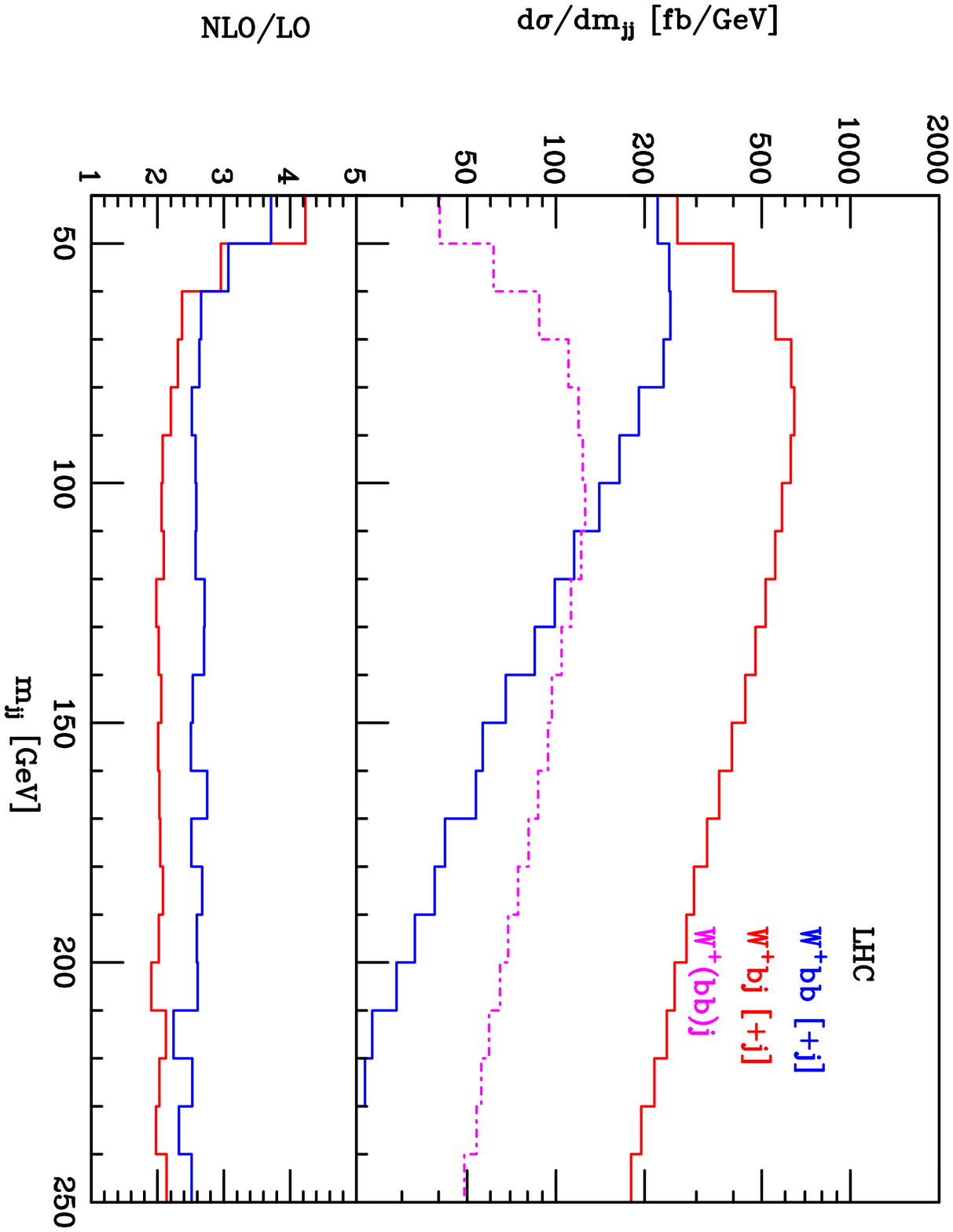}
\end{sideways}
\vspace*{-.8cm}
\end{center}
\caption{Same as Fig.~\ref{fig:highpt-lhc}, but for the di-jet
invariant mass.  For $Wbb[+j]$, the invariant mass is that of the
two $b$ jets; for $Wbj[+j]$, the invariant mass is that of the $b$
jet and the other jet with the higher $p_T$.}
\label{fig:massinv-lhc}
\end{figure}

We show in Figs.~\ref{fig:highpt-lhc},\ref{fig:lepton-lhc},
\ref{fig:massinv-lhc} the same distributions at the LHC, but this
time for the inclusive cross sections, rather than exclusive.  At
the LHC, the process $Wbj+X$ is dominant, while $Wbb+X$ and $W(bb)j$
are comparable.  There is little change in shape of the
distributions at NLO, with the exception of the $p_T$ spectrum of
the highest $p_T$ jet, which is enhanced at high $p_T$ for $Wb\bar
b+X$.  The di-jet invariant mass distribution also shows an
enhancement at low values of the invariant mass, but not at values
relevant for the Higgs search.


The code MCFM is publicly available \cite{MCFM}, and one can
generate any desired distribution.  We hope the calculation outlined
in this paper is useful to better understand the backgrounds to
single-top production, Higgs production, and signals of new physics
at the Tevatron and the LHC.

\section*{Acknowledgments}

\indent\indent We are grateful for conversations and correspondence
with C.~Ciobanu, J.~Ellison, A.~Heinson, T.~Junk, T.~Liss,
T.~McElmurry, F.~Olness, K.~Pitts, and R.~Schwienhorst. J.~C.\ and
S.~W.\ thank the Aspen Center for Physics for hospitality.  This
work was supported in part by the U.~S.~Department of Energy under
contracts Nos.~DE-AC02-76CH03000 and DE-FG02-91ER40677.


\end{document}